# A HEURISTIC APPROACH TO THE RELATIVISTIC ANGULAR MOMENTUM

Sebastiano Tosto

Enea CRE Casaccia, via Anguillarese 301     00060 Roma, Italy

ABSTRACT

The relativistic angular momentum is introduced as an extension of the non-relativistic analysis of allowed states in the phase space for a quantum particle. The paper shows the conceptual basis of the approach. An interesting feature of the present point of view is that the indistinguishability of identical particles and the Pauli principle are found as corollaries.

1 Introduction.

Two papers have been recently published concerning the analysis of states allowed to the particles in the phase space to calculate the energy levels of many electron atoms /1/ and diatomic molecules /2/. The properties of the quantum angular momentum were also found as a straightforward consequence of the basic assumptions described in these papers. However the non-relativistic character of this approach did not enable to infer any information about the spin angular momentum of electrons. The relativistic quantum theory /3,4/ has introduced the spin as an intrinsic property of particles; perturbation calculations of the energy levels including the spin-orbit and spin-spin couplings are widely reported in literature, see e.g. /5/. On this respect, however, it is interesting to inquire whether the existence of spin can be also inferred through an extension of the non-relativistic approach to the angular momentum with the help of Lorentz transformation. The present paper aims just to discuss the fundamental ideas leading to the concept itself of spin through the analysis of the phase space; from a conceptual point of view this point appears more attracting than the development of a further more or less approximate numerical calculation algorithm including the spin effects of specific interest for the solution of some particular quantum problem. The paper is organized as follows:

-section 2 summarizes for clarity the physical background of the present approach;

-section 3 concerns some key-points about the relativistic angular momemtum;

-section 4 introduces the approach to the angular momentum of quantum particles;

- the results obtained are discussed in section 5;

-the conclusions are reported in section 6.

2 Physical background of the non-relativistic approach.

The state of a quantum particle is described by a wave function $\psi$ defined as the solution of the appropriate partial differential wave equation subjected to the pertinent boundary conditions. Our knowledge about the system rests in general just on the possibility to write down this wave equation and find its solution. By consequence, the deterministic concepts of position or trajectory of classical physics are replaced by the probability density, given by $\psi\psi^*$ if $\int \psi\psi^* dV = 1$, of finding the particle in a given region of space. The papers /1,2/ are based on a different, more agnostic, physical idea according which one renounces "*a priori*" even to the probability character of our knowledge. Rather, since the beginning one assumes the total uncertainty about both position and momentum of the particle. For sake of clarity, this idea is explained with the help of two examples, the first of which concerns just the angular momentum $\vec{M} = \vec{r} \times \vec{P}$ of a particle having a momentum $\vec{P}$. According to the initial assumption, the current position of the particle is regarded as completely unknown; the only information available about $\vec{r}$ is that $0 < |\vec{r}| \leq |\Delta \vec{r}|$, i.e. $|\Delta \vec{r}|$ has the meaning of quantum uncertainty range for $|\vec{r}|$. The same holds also for the momentum of the particle; $\vec{P}$ is assumed unknown as well, while being $0 < |\vec{P}| \leq |\Delta\vec{P}|$, i.e. $|\vec{P}|$ must fall within its uncertainty range $|\Delta\vec{P}|$. No hypotheses are necessary nor about $|\Delta\vec{r}|$ neither about $|\Delta\vec{P}|$; this follows because no hypotheses have been made about $|\vec{r}|$ and $|\vec{P}|$ themselves and because the uncertainty ranges are defined merely as those including all the possible current values of the dynamical variables. Even in lack of any detailed information about the motion of the particle, the total number of states $l$ consistent with the angular motion of the particle can be calculated just considering the widths $|\Delta\vec{r}|$ and $|\Delta\vec{P}|$ of the uncertainty ranges; clearly, in fact $l$ is not related to the actual, randomly changing position and momentum of the particle, but rather to the total ranges allowed to the dynamical variables describing its angular motion. Once having renounced "*a priori*" to any knowledge about the state of motion of the particle, $l$ is the only information available. The maximum values allowed to $|\vec{r}|$ and $|\vec{P}|$ are just $|\Delta\vec{r}|$ and $|\Delta\vec{P}|$ in agreement with the definition above. Therefore $l$ can be calculated simply considering the particular case $|\vec{r}| \equiv |\Delta\vec{r}|$ and $|\vec{P}| \equiv |\Delta\vec{P}|$ where the dynamical variables assume their maximum values, in practice, replacing $|\vec{r}|$ with $|\Delta\vec{r}|$ and $|\vec{P}|$ with $|\Delta\vec{P}|$ in $M = |\vec{M}|$. Let us consider therefore $M = M(|\vec{r}|, |\vec{P}|)$ in the particular case

$M = M(|\Delta \vec{r}|, |\Delta \vec{P}|)$ in order to find $M = M(l)$. The starting point to calculate $l$ is the component $M_n = \vec{M} \cdot \vec{n}$ of $\vec{M}$ along an arbitrary direction defined by the unit vector $\vec{n}$. The classical expression $M_n = (\vec{r} \times \vec{P}) \cdot \vec{n}$ is replaced by $\Delta M_n = (\Delta \vec{r} \times \Delta \vec{P}) \cdot \vec{n}$. Rewriting $\Delta M_n$ identically as $\Delta M_n = (\vec{n} \times \Delta \vec{r}) \cdot \Delta \vec{P}$ one obtains $\Delta M_n = \Delta \vec{\xi} \cdot \Delta \vec{P}$, being $\Delta \vec{\xi} = \vec{n} \times \Delta \vec{r}$. If $\Delta \vec{P}$ and $\Delta \vec{\xi}$ are orthogonal then $\Delta M_n = 0$; else, by writing the scalar product $\Delta \vec{\xi} \cdot \Delta \vec{P}$ as $(\Delta \vec{P} \cdot \Delta \vec{\xi} / \Delta \xi) \Delta \xi$, where $\Delta \xi$ is the modulus of $\Delta \vec{\xi}$, the component $\pm \Delta P_\xi = \Delta \vec{P} \cdot \Delta \vec{\xi} / \Delta \xi$ of $\Delta \vec{P}$ along $\Delta \vec{\xi}$ gives $\Delta M_n = \pm \Delta \xi \Delta P_\xi$. In turn, this latter equation gives $\Delta M_n = \pm l \hbar$ where $l = 1, 2 \cdots$. These results are summarized as

$$\Delta M_n = \pm l \hbar \qquad \text{where} \qquad l = 0, 1, 2, \cdots \qquad 2,1$$

Ref /1/ shows also: (i) the impossibility to know simultaneously all the components of angular momentum and (ii) that $\Delta M^2 = l(l+1)\hbar^2$ also follows from eq 2,1. $l$ is arbitrary because no hypotheses has been made on $\vec{r}$, $\vec{P}$, $\vec{n}$ nor on $|\Delta \vec{r}|$ and $|\Delta \vec{P}|$. As expected, the quantities of relevant physical interest for the properties of the angular momentum are $|\Delta \vec{r}|$ and $|\Delta \vec{P}|$; in fact, the key to introduce the quantization of $M_n$ was the step $M_n(|\vec{r}|, |\vec{P}|) \to M_n(|\Delta \vec{r}|, |\Delta \vec{P}|)$ instead of the replacement of the dynamical variables with the respective operators: by consequence $l$ is now a number of states and not the mathematical result of any boundary condition imposed to the wavefunction describing the particle motion. The notation $\Delta M_n$ is therefore not merely formal: it means physically that exists a range of discrete values for $M_n$ defined by the possible numbers $l$ of quantum states allowed to the particle. The classical uncertainty imposed by considering $|\vec{r}|$ and $|\vec{P}|$ as randomly changing within $|\Delta\vec{r}|$ and $|\Delta\vec{P}|$ results expressed therefore from the quantum point of view through the range of arbitrary values possible for $l$. This simple example explains why in the present approach the actual values of the dynamical variables are of no interest. Moreover, owing to the very general character of the ideas just introduced, it appears since now that the quantum properties of particles inferred through the total uncertainty are consistent with that obtained by solving the wave equation. In effect, the steps shortly sketched above have been extended in /1/ to the cases of the particle in a box, to the harmonic oscillator and to the energy levels of multielectron atoms, in which case the results calculated are in a very good agreement with the experimental data; in the same way were also calculated the fundamental vibrational frequency, bond length and binding energy of homonuclear and etheronuclear diatomic molecules treated as anharmonic oscillators /2/. It is also instructive for the purposes of the present paper to summarize only the case of hydrogenlike atoms to show how this approach applies to calculate the electron energy levels. The classical energy equation of an electron in the field of a nucleus with charge $Z$ in the reference system fixed on the center of mass is

$$E = E_B + \frac{P_\rho^2}{2\mu} + \frac{M^2}{2\mu\rho^2} - \frac{Ze^2}{\rho} \qquad E = E(\rho, P_\rho, M^2)$$

where $\mu$ is the reduced mass, $\rho$ and $P_\rho$ the moduli of radial distance and momentum. The basic assumption that the electron is completely delocalized around the nucleus requires to replace $\rho$ and $P_\rho$, now to be considered unknown, with the respective uncertainty ranges $\Delta\rho$ and $\Delta P_\rho$. The same holds also for $\vec{M}$ to be replaced by $\Delta\vec{M}$ as discussed above. Again, no hypotheses are necessary about $\rho$ and $P_\rho$ and then about $\Delta\rho$ and $\Delta P_\rho$. At this point the only information available are the numbers of quantum states $n$ and $l$ related to the radial and angular motion of the electron. Nevertheless, the electron energy levels $E(n,l)$ can be effectively obtained by replacing $\rho$, $P_\rho$ and $M^2$ with $\Delta\rho$, $\Delta P_\rho$ and $\Delta M^2$ in $E = E(\rho, P_\rho, M^2)$, exactly as discussed above. The classical energy equation is rewritten as

$$E^* = E_B + \frac{\Delta P_\rho^2}{2\mu} + \frac{\Delta M^2}{2\mu\Delta\rho^2} - \frac{Ze^2}{\Delta\rho} \qquad E^* = E^*\left(\rho \equiv \Delta\rho, P_\rho \equiv \Delta P_\rho, M^2 \equiv \Delta M^2\right)$$

$n$ is given by $n = (2\Delta\rho\Delta P_\rho / \hbar)/2$. The factor 2 within parenthesis accounts for the possible states of spin of the electron. The factor ½ is due to the fact that really $P_\rho^2$ is consistent with two possible values $\pm P_\rho$ of the radial component of the momentum corresponding to the inwards and outwards motion of the electron with respect to the nucleus. By consequence, being the uncertainty range $\Delta P_\rho$ clearly the same in both cases, the calculation of $n$ as $2\Delta\rho\Delta P_\rho / \hbar$ would mean to count separately two different situations both certainly possible for the electron but really corresponding to the same quantum state. These situations are in fact physically indistinguishable because of the total uncertainty assumed "*a priori*" about the motion of the electron; therefore the factor ½ avoids to count twice a given quantum state. Again, $n$ and $l$ take in principle any integer values because the uncertainty ranges $\Delta\rho$ and $\Delta P_\rho$ include arbitrary values of ρ and $P_\rho$ and then are arbitrary themselves. Replacing $\Delta P_\rho$ with $n\hbar/\Delta\rho$ and $\Delta M^2$ with $(l+1)l\hbar^2$ in $E^*$ the result is

$$E^* = E_B + \frac{n^2\hbar^2}{2\mu\Delta\rho^2} + \frac{l(l+1)\hbar^2}{2\mu\Delta\rho^2} - \frac{Ze^2}{\Delta\rho} \qquad n = 1,2,3,... \qquad l = 0,1,2,...$$

With elementary manipulations this equation reads

$$E^* = E_B + \frac{1}{2\mu}\left(\frac{n\hbar}{\Delta\rho} - \frac{Ze^2\mu}{n\hbar}\right)^2 + \frac{(l+1)l\hbar^2}{2\mu\Delta\rho^2} - \frac{Z^2e^4\mu}{2n^2\hbar^2}$$

It is possible to minimize $E^*$ putting equal to zero the quadratic term within parenthesis, certainly positive; being $E = \min(E^*)$ the result is

$$\Delta\rho = \frac{n^2\hbar^2}{Ze^2\mu} \qquad E = E_B + \frac{(l+1)l\hbar^2}{2\mu\Delta\rho^2} - \frac{Z^2e^4\mu}{2n^2\hbar^2}$$

Then the total quantum energy $E(n,l)$ of the hydrogenlike atom results as a sum of three terms: (i) the kinetic energy of the center of mass of the atom considered as a whole, (ii) the quantized rotational energy of a system having a reduced mass $\mu$ with the electron at a distance $\Delta\rho$ from the nucleus and (iii) a negative term to be necessarily identified as the non-relativistic binding energy $E_{el}$ of the electron. The values allowed to $l$ must fulfill the condition $l \leq n$. Let us rewrite in fact $E$ in a reference system where the center of mass is at rest, $E_B = 0$, utilizing the expression of $\Delta\rho$ just found

$$E = \left[\frac{(l+1)l}{n^2} - 1\right]\frac{Z^2 e^4 \mu}{2n^2 \hbar^2}$$

If $l \geq n$ then the total energy $E$ would result $\geq 0$, i.e. the hydrogen atom would not be in a bound state. Then, the stability of hydrogenlike atom requires an upper value $n-1$ for $l$. Hence it is possible to write $n = n_o + l + 1$, where $n_o$ is of course still an integer. One finds

$$E_{el} = -\frac{Z^2 e^4 \mu}{2(n_o + l + 1)^2} \qquad n_o = 1,2,3,... \qquad l = 0,1,2,... \qquad 2,2$$

Therefore, all the possible terms expected for the non-relativistic energy are found in a straightfoward and elementary way. The conceptual connection between the classical and quantum energies $E(\rho, P_\rho, M^2) \to E(\Delta\rho, \Delta P_\rho, \Delta M^2) \to E(n,l)$ is the same as that discussed for the angular momentum: it is obtained again by replacing the dynamical variables of the classical hamiltonian with the respective uncertainty ranges rather than with the respective operators. This position, merely aimed to count the quantum states allowed to the electron, explains why the calculations for the cases shortly outlined here proceed through simple algebraic manipulations rather than by solving the appropriate wave equations. Moreover, these results also explain why the current values of the dynamical variables can not appear in the final expressions of the angular momentum and energy levels of hydrogenlike atoms even when obtained from the respective wave equations. Ref /1/ shows that these ideas hold also to calculate in straightforward way the energy levels of multielectron atoms. Also now, $n$ and $l$ are not due to any quantization condition on the motion of the electron but rather, owing to the initial total uncertainty about its dynamical variables, represent numbers of allowed states. This fact has an important consequence: any reference to a specific electron is conceptually lost since the beginning. Instead of being properties of the electron, $n$ and $l$ are pertinent to the ranges $\Delta\rho$ and $\Delta P_\rho$ where *any* electron could be found; in effect, the uncertainty principle itself concerns a given number of states regardless of the kind of particle itself or its actual dynamical variables. In turn, it means that     and the energy levels of multielectron atoms are expressed in principle without any concern to which electron in particular belongs to a given state. Hence, this approach implies necessarily the indistinguishability of identical particles; in fact, it is physically meanigless any possibility to distinguish particles whose dynamical variables have been ignored conceptually, since the beginning and not as a sort of numerical approximation merely aimed to simplify some calculation. The indistinguishability is found now as a corollary rather than being introduced as a postulate.

It is clear at this point the interest to check whether or not the ideas so far discussed hold again in the frame of a relativistic approach. This check is important in general, because the consistency with the relativity is certainly a necessary conceptual requirement for any physical theory. Moreover, owing to their non-relativistic character, the papers /1,2/ were of course unable to explain why the numbers of states $n$ and $l$ should fulfil the Pauli principle in order to calculate correctly the electron energy levels of multielectron atoms and diatomic molecules. The next sections aim to clarify this point just introducing the requirements of relativity into the present approach.

3 The relativistic angular momentum.

The ideas introduced in the previous section must be modified when taking into account the basic requirements of relativity. It is known in general that in relativity the angular momentum of a system of particles is an antisymmetric four tensor built of two three-vectors /6/: $\vec{M} = \Sigma_j (\vec{r} \times \vec{P})$ and $\vec{M}_4 = ic\Sigma_j (t\vec{P} - \varepsilon \vec{r}/c^2)$, where the summation is extended to the number $j$ of particles of the system. $\vec{M}_4$ is defined by the center of inertia of the system of particles. In the following we concern the case of a free particle, i.e. $j = 1$. Let us consider to this purpose two inertial reference systems $R$ and $R'$ moving with relative constant velocity $\vec{V}$ and a quantum particle whose linear momentum and proper distance from the origin are $\vec{P}$ and $\vec{r}$ respectively in $R$. Then, the proper length of the vector $\vec{r}$ appears contracted to a value $\vec{r}'$ for an observer in $R'$. Let us assume without loss of generality that the origins of the reference systems $R$ and $R'$ coincide at the time $t=0$ and are displaced by a distance $\vec{V}\delta t$ after a time range $\delta t$, being both $t$ and $\delta t$ defined in $R$. For an observer in $R'$ /7/

$$\vec{r}' = \vec{r} - \lambda^* \vec{V} \quad \text{3,1a} \qquad \text{where} \qquad \lambda^* = \frac{\vec{r}\cdot\vec{V}}{V^2} - \frac{\dfrac{\vec{r}\cdot\vec{V}}{V^2} - \delta t}{\sqrt{1 - \dfrac{V^2}{c^2}}} \qquad \text{3,1b}$$

Analogous considerations hold to relate the linear momentum $\vec{P}'$ in $R'$ and $\vec{P}$ in $R$. A procedure similar to that followed to derive eq 3,1b, sketched in appendix A, gives

$$\vec{P}' = \vec{P} - \mu\vec{V} \quad \text{3,2a} \qquad \text{where} \qquad \mu = \frac{\vec{P}\cdot\vec{V}}{V^2} - \frac{\dfrac{\vec{P}\cdot\vec{V}}{V^2} - \dfrac{\varepsilon}{c^2}}{\sqrt{1 - \dfrac{V^2}{c^2}}} \qquad \text{3,2b}$$

where $\varepsilon$ is the energy of the particle. Then $\vec{M}' = \vec{r}' \times \vec{P}'$ in $'$ is calculated as a function of $\vec{M} = \vec{r} \times \vec{P}$ in $R$ through eqs 3,1 and 3,2

$$\vec{M}' = \vec{M} - \lambda^*\vec{V}\times\vec{P} - \mu\vec{r}\times\vec{V} \qquad 3,3$$

Elementary manipulations, shortly sketched in appendix B, show that eq 3,3 is identical to that obtained directly from the general theory of Lorentz transformations of 4-tensors

$$\vec{M}' = \frac{1}{\sqrt{1-V^2/c^2}}\left[\vec{M} + \frac{\vec{V}}{V^2}(\vec{V}\cdot\vec{M})(\sqrt{1-V^2/c^2}-1) - \vec{V}\times\left(t\vec{P} - \frac{\varepsilon}{c^2}\vec{r}\right)\right]$$

The next section shows how the transformation properties of eq 3,3 enable to modify eq 2,1, thus introducing the quantized relativistic angular momentum.

4 Introduction to the relativistic angular momentum.

Summing and subtracting the vector $\vec{V}\delta t$ at right hand side of eq 3,1a one obtains

$$\vec{r}' = \vec{r}_G - \lambda\vec{V} \qquad 4,1$$

being

$$\vec{r}_G = \vec{r} - \vec{V}\delta t \qquad 4,2a \qquad \text{and} \qquad \lambda = \frac{\vec{r}\cdot\vec{V}}{V^2} - \delta t - \frac{\dfrac{\vec{r}\cdot\vec{V}}{V^2} - \delta t}{\sqrt{1-\dfrac{V^2}{c^2}}} \qquad 4,2b$$

Eq 4,2b confirms that $\vec{r}_G$ is a galileian transformation of $\vec{r}$; in effect, if $c$ tends to infinity then $\lambda \to 0$, so that $\vec{r}'$ tends to $\vec{r}_G$. In conclusion, $\vec{r}'$ is equivalent to a galileian transformation $\vec{r}_G$ of $\vec{r}$ less a relativistic correction $\lambda\vec{V}$ calculated with the help of eq 4,2b. Hence, the moduli of $\vec{r}$ and $\vec{r}_G$ must be equal

$$|\vec{r}_G| = |\vec{r}| \qquad 4,3$$

Then eqs 4,2a and 4,3 give

$$2\vec{r}\cdot\vec{V}\delta t = (\vec{V}\delta t)^2 \qquad 4,4$$

i.e.

$$\frac{\vec{r}\cdot\vec{V}}{V^2} = \frac{1}{2}\delta t \qquad 4,5$$

Then it is possible to rewrite $\vec{M}'$ as a function of $\vec{r}_G$ utilizing eqs 3,2a and 4,1

$$\vec{M}' = \vec{r}' \times \vec{P}' = (\vec{r}_G - \lambda \vec{V}) \times (\vec{P} - \mu \vec{V})$$

where $\lambda$ and $\mu$ are given by eqs 4,2b and 3,2b respectively. Then we obtain

$$\vec{M}' = \vec{L} + \vec{S} \qquad 4,6$$

where

$$\vec{L} = \vec{r}_G \times (\vec{P} - \mu \vec{V}) \quad 4,7a \qquad \text{and} \qquad \vec{S} = -\lambda \vec{V} \times \vec{P} \quad 4,7b$$

The relativistic linear momentum $\vec{P}$ of a particle having rest mass $m$ and velocity $\vec{v}$ is

$$\vec{P} = \frac{m\vec{v}}{\sqrt{1 - \frac{v^2}{c^2}}} \qquad 4,8$$

In the case of a free particle not subjected to any interactions one assumes that the velocity $\vec{v}$ is a constant. Replacing eq 4,5 into eq 4,2b, $\vec{S}$ of eq 4,7b becomes

$$\vec{S} = \frac{1}{2}\left[\vec{V}\delta t - \frac{\vec{V}\delta t}{\sqrt{1-\frac{V^2}{c^2}}}\right] \times \frac{m\vec{v}}{\sqrt{1-\frac{v^2}{c^2}}} \qquad 4,9$$

If $c$ is put equal to infinity the component $\vec{L}$ of $\vec{M}'$, eq 4,7a, takes the classical form $\vec{L} = \vec{r}_G \times m\vec{v}$ because $\mu$ vanishes according to eq 3,2b, whilst the component $\vec{S}$ vanishes. The vector $\vec{S}$ is therefore a relativistic correction to $\vec{\ }$. Let us recall now that $\delta t$ is a time range in $R$. The time range $\delta t'$ for an observer in $R'$ corresponding to $\delta t$ is defined by

$$\delta t = \delta t'\sqrt{1-\frac{V^2}{c^2}} \qquad 4,10$$

Then, eq 4,9 gives

$$\vec{S} = \frac{1}{2}(\delta t - \delta t')\vec{V} \times \frac{m\vec{v}}{\sqrt{1-\frac{v^2}{c^2}}} \qquad 4,11$$

It appears also in eq 4,11 that $\vec{S}$ vanishes if $c \to \infty$, because in this case $\delta t' \to \delta t$. It is now formally possible to exchange the vectors $\vec{v}/\sqrt{1-v^2/c^2}$ and $\vec{V}$ in the cross product of eq 4,11, thus obtaining

$$\vec{S} = \frac{1}{2} \frac{\delta \vec{\varphi}}{\sqrt{1-\frac{v^2}{c^2}}} \times \vec{P}_V \qquad 4,12$$

where

$$\delta \vec{\varphi} = (\delta t' - \delta t)\vec{v} \qquad \text{and} \qquad \vec{P}_V = m\vec{V} \qquad 4,13$$

Also, it is possible to recognize in eq 4,12 the vector $\delta \vec{\varphi}_v$ defined as follows

$$\delta \vec{\varphi}_v = \frac{\delta \vec{\varphi}}{\sqrt{1-\frac{v^2}{c^2}}} \qquad 4,14$$

In turn, $\delta \vec{\varphi}$ can be regarded as the Lorentz contraction of the proper length $\delta \vec{\varphi}_v$ defined in a reference system moving with a constant velocity $\vec{v}$, i.e. solidal with the particle itself. Then $\vec{S}$ of eq 4,12 is expressed as a function of $\delta \vec{\varphi}_v$ through eq 4,14

$$\vec{S} = \frac{1}{2}\delta\vec{\varphi}_v \times \vec{P}_V \qquad 4,15$$

To summarize: $\vec{M}'$ in the reference system $R'$ is expressed through a component $\vec{L}$, given by eq 4,7a, plus a relativistic component $\vec{S}$, given by eq 4,15 and due to the contraction $\vec{r}'$ of the proper length $\vec{r}$; in fact $\vec{S}$ has been initially introduced because in eq 4,1 the simple galileian transformation $\vec{r}_G$ of $\vec{r}$ has been replaced by $\vec{r}_G - \lambda\vec{V}$. It is easy to realize that the angular momentum $\vec{S}$ does not depend on the state of motion of the particle, but rather it is an intrinsic property of the particle itself. In fact:

(i) according to eq 4,13, $\vec{P}_V$ is related merely to the drift speed $\vec{V}$ of the reference system $R$ with respect to $R'$. It is confirmed by the fact that $\vec{S}$ of eq 4,7b could have been identically written as $\vec{S} = -\lambda\vec{V} \times (\vec{P} + \vec{P}_V)$, i.e. replacing $\vec{P}$ with $\vec{P} + \vec{P}_V$ without affecting $\vec{M}'$.

(ii) $\delta\vec{\varphi}_v$ is, by its own definition, a proper length solidal to the particle, therefore an internal degree of freedom of the particle rather than a vector characterized in some way by its state of motion.

(iii) for a free particle $\vec{P} = \varepsilon\vec{v}/c^2$, therefore $\vec{M}_4 = (\vec{v}t - \vec{r})\varepsilon/c$. The conservation law of angular momentum requires that $\vec{M}_4$ of a free particle be a constant. Being $\varepsilon$ constant too, one infers /6/ that $\vec{v}t - \vec{r} = const$; this merely means that the vector $\vec{r}$ moves with a velocity $\vec{v}$.

The results so far obtained, although deduced assuming initially a constant velocity $\vec{v}$ of the particle in eq 4,8, hold in general: in fact, the meaning of $\vec{S}$ does not change even for a different state of motion of the particle, being lost any reference to the particular value of $\vec{v}$. Rather, $\vec{S}$ is here merely a consequence of Lorentz transformations for the inertial references $R$ and $R'$. At this point we assume that the basic ideas to derive the non-relativistic quantum angular momentum still hold here and can be applied to handle also the cross product $\delta\vec{\varphi}_V \times \vec{P}_V$ of eq 4,15 in the same way as shortly sketched in paragraph 2 for the non-relativistic vectors. In other words, also these relativistic dynamical variables are treated as quantities whose allowed values fall within ranges having the physical meaning of quantum uncertainties. Then, the procedure summarized in section 2 shows immediately that the component of the cross product $\delta\vec{\varphi}_V \times \vec{P}_V$ along an arbitrary direction defined by $\vec{n}$ is equal to $\pm s\hbar$, being $s$ an integer including zero. In conclusion, the component of $\vec{S}$ along $\vec{n}$ is

$$S_z = \pm \frac{1}{2}s\hbar \qquad s = 0,1,2,... \qquad 4,16$$

The component of the total angular momentum of the particle is then, thanks to eq 4,6

$$M_z = \pm l\hbar \pm \frac{1}{2}s\hbar \qquad 4,17$$

The properties of the additional angular momentum $\vec{S}$, for instance the impossibility to know simultaneously its x, y and z components, are found with the same procedure followed in /1/ for $\vec{L}$.

5. Discussion

The results obtained in the previous paragraph have shown that the Lorentz transformations are conditions enough and necessary for the existence of an angular momentum quantum number $s$ additional with respect to $l$. All the considerations so far carried out require only that $c$ is finite: any comparison between the velocity of the quantum particle with respect to $c$ is in principle irrelevant. In fact, the above discussion has shown that only $c = \infty$ means in any case $\vec{S} = 0$ and $\vec{L} = \vec{r}_G \times m\vec{v}$, regardless of the value of $\vec{v}$, thus obtaining only the non-relativistic component $l\hbar$ of the angular momentum. This conclusion follows when both coniugate dynamical variables fulfil the Lorentz transformations. In particular, it is essential that: (i) the relativistic dynamical variables be regarded as randomly changing within the respective quantum uncertainties and (ii) the number of allowed states in the relativistic phase space be calculated through these quantum uncertainties. In effect, the idea to extend the approach shortly sketched in section 2 to the relativistic case is reasonable because the Lorentz transformations change merely the analytical expressions of momenta and coniugate lengths but not their own physical meaning. On this respect, it is interesting the fact that the analysis of states in the relativistic phase space allows to describe also a form of angular momentum that, strictly speaking, is an intrinsic property of the quantum particle rather than a true kinematical property. An important consequence of these results is inferred considering an arbitrary number $N$ of identical particles whose state is described by positions and linear momenta falling within proper uncertainty ranges of the phase space. As shown above, the number of states allowed for the angular momentum of each particle is given now by a number $l$, depending on the motion of the particle, and by a number $s$, characteristic of the particle itself regardless of its state of motion. Let us consider now separately the cases where $s$ is either even, including zero, or odd. It is immediate to realize that in the former case a given number of states $n_{tot}$ in the phase space can be defined regardless of the actual number of particles: in fact, being $l$ arbitrary and both $s$ and $s/2$ integers, a situation where, say, $M_z = 50\hbar$ could result because, for instance, $l + s/2 = 50$ for one particle or $l + s/2 = 2$ for each one of 25 particles. In other words, it means that any increasing of the number of particles in a system does not lead to a situation in the phase space physically distinguishable from the previous one: hence the number of particles consistent with a given quantum state is in principle arbitrary, i.e. an arbitrary number of $s$ even particles can be found in a given quantum state. The conclusion is completely different when considering a system of particles characterized by $s$ odd; in this case the properties of the phase space are not longer indistinguishable with respect to the addition of particles because now the respective values of $M_z$ jump from ...integer, half-integer, integer ... values upon addition of each further particle: any change of the number of particles necessarily gives a total component of $M_z$ different from the previous one. In other words, any further odd $s$ particle added to the system is really described by a new quantum state distinguishable from those already existing, then necessarily a quantum state different from that of the other particles.

Two remarks are necessary at this point.

The first remark concerns the question posed in section 2, i.e. why the numbers of states $n$ and $l$ appearing in eqs 2,1, 2,2 and in the expressions of many-electron atoms and diatomic molecules reported in /1,2/ must fulfil the Pauli principle in order that the results be effectively correct. The simple answer is that $n$, $l$ and    must necessarily increase as long as increases the number of electrons simply because each electron must be in its own quantum state; then $n$ and $l$ can remain the same only if the further quantum state necessary for a new electron can be accounted for by both possibilities for $M_z$ of eq 4,17.

The second remark concerns the physical meaning of $\vec{S}$ itself. It is evident at this point the connection of the present results with the behaviour of bosons and fermions as concerns the Pauli principle. According to this latter no more than one fermion can occupy the *same* quantum state; the present conclusion was that the quantum state of each *s* odd particle is necessarily different from that of other particles. These statements are only formally different but clearly equivalent in principle. This supports the idea that $\vec{S}$ is just the spin angular moment, in which case the discussion above shows that the Pauli principle is really a corollary according to the present point of view. On this respect, it is significant to remind that also the indistinguishability of identical quantum particles follows in the present approach as a corollary for the reasons discussed in detail in /1/ and shortly sketched in section 2.

6 Conclusion.

The present approach has described the angular momentum as a straightforward consequence of the quantum uncertainty in the space phase of conjugate relativistic dynamical variables. It is remarkable that the two fundamental statements of quantum mechanichs, i.e. the indistingushability of identical particles and the Pauli principle, are obtained as corollaries rather than being postulates imposed "*a priori*". It appears therefore that, really, information is gained while renouncing to any kind of information about the dynamical variables of the particles themselves. This approach is the first step towards to the description of the relativistic many electron atom, taking into account not only the coulombian repulsion but also the coupling between the spin and orbital angular momenta of the electrons. Work is in advanced progress on this subject.

# APPENDIX A

In the case where two inertial reference systems $R$ and $R'$ move with relative velocity $V_x$ oriented along the x-axis the transformations of lengths and momenta are /7/

$$P'_x = \frac{P_x - \varepsilon \frac{V_x}{c^2}}{\sqrt{1 - \frac{V_x^2}{c^2}}}$$

being $\varepsilon$ the energy of the particle. A general vector formula of transformation for the momentum is easily found assuming that $P_x = \vec{P} \cdot \vec{V} / |\vec{V}|$ and $P'_x = \vec{P}' \cdot \vec{V} / |\vec{V}|$. Let us insert $P'_x$ and $P_x$ in the above formula and take into account that the component of the momentum normal to the velocity is not affected by the Lorentz contraction, i.e.

$$\vec{P}' - \frac{\vec{P}' \cdot \vec{V}}{V^2} \vec{V} = \vec{P} - \frac{\vec{P} \cdot \vec{V}}{V^2} \vec{V}$$

Utilizing also this latter result we obtain with elementary manipulations

$$\vec{P}' = \vec{P} - \left[ \frac{\vec{P} \cdot \vec{V}}{V^2} - \frac{\frac{\vec{P} \cdot \vec{V}}{V^2} - \frac{\varepsilon}{c}}{\sqrt{1 - \frac{V^2}{c^2}}} \right] \vec{V}$$

APPENDIX B

Utilizing eqs 3,1b and 3,2b for $\lambda^*$ and $\mu$, eq 3,3 reads

$$\vec{M}' = \vec{M} + \left[\frac{\vec{r}\cdot\vec{V}}{V^2}\left(1-\frac{1}{\beta}\right)+\frac{t}{\beta}\right]\vec{P}\times\vec{V} - \left[\frac{\vec{P}\cdot\vec{V}}{V^2}\left(1-\frac{1}{\beta}\right)+\frac{\varepsilon/c^2}{\beta}\right]\vec{r}\times\vec{V}$$

where $\quad \beta = \sqrt{1-\frac{V^2}{c^2}}$

Collecting $\beta$ this equation gives

$$\vec{M}' = \frac{1}{\beta}\left\{\left[\vec{M}+\frac{\vec{r}\cdot\vec{V}}{V^2}\vec{P}\times\vec{V}-\frac{\vec{P}\cdot\vec{V}}{V^2}\vec{r}\times\vec{V}\right]\beta - \frac{\vec{r}\cdot\vec{V}}{V^2}\vec{P}\times\vec{V}+\frac{\vec{P}\cdot\vec{V}}{V^2}\vec{r}\times\vec{V}+\left(t\vec{P}-\frac{\varepsilon}{c^2}\vec{r}\right)\times\vec{V}\right\}$$

On the other hand, it is immediate to show that

$$\vec{M}+\frac{\vec{r}\cdot\vec{V}}{V^2}\vec{P}\times\vec{V}-\frac{\vec{P}\cdot\vec{V}}{V^2}\vec{r}\times\vec{V} = \frac{\vec{V}}{V^2}\vec{M}\cdot\vec{V} \qquad \text{B1}$$

In fact, recalling that $\vec{u}_1\times(\vec{u}_2\times\vec{u}_3) = (\vec{u}_1\cdot\vec{u}_3)\vec{u}_2 - (\vec{u}_1\cdot\vec{u}_2)\vec{u}_3$ eq B1 reads

$$\left[ (\vec{r} \cdot \vec{V})\vec{P} - (\vec{P} \cdot \vec{V})\vec{r} + \vec{V} \times \vec{M} \right] \times \vec{V} = 0$$

Furthermore, being $\vec{M} = \vec{r} \times \vec{P}$, then $\vec{V} \times (\vec{r} \times \vec{P}) = (\vec{V} \cdot \vec{P})\vec{r} - (\vec{V} \cdot \vec{r})\vec{P}$ so that effectively the left hand side of the last equation vanishes, thus proving eq B1. On the other hand, replacing eq B1 into eq 3,3 one immediately finds just the transformation equation of the 4-tensor angular momentum reported at the end of section 3.

REFERENCES

1 S. Tosto, Il Nuovo Cimento B, vol 111, n. 3, 1996, pp. 193-215

2 S. Tosto, Il Nuovo Cimento D, vol 18, n. 13, 1996, pp. 1363-1394

3 J. Aharony, "The special theory of relativity", Clarendon Press, Oxford, 1959

4 J. L. Synge, "Relativity: the special theory", North-Holland Publishing Company, Amsterdam, 1973, pp 102-110

5 J. P. Desclaux, "Relativistic Multiconfiguration Dirac-Fock package", in the book "Methods and Techniques in Computational Chemistry: METECC-94", vol A: small systems, E. Clementi Ed., STEF, Cagliari, 1993

6 L. Landau and E. Lifchitz, "Théorie du Champ", Editions MIR, Moscou, 1966, p 56

7 A. S. Kompaneiets, "Theoretical Physics", MIR Publishers, Moskow, 1966, p 219